\documentclass[12pt]{article}

\usepackage{graphicx}
\begin{document}

\begin{center}
{\bf New model of 4D Einstein$-$Gauss$-$Bonnet gravity coupled with nonlinear electrodynamics} \\
\vspace{5mm} S. I. Kruglov
\footnote{E-mail: serguei.krouglov@utoronto.ca}
\underline{}
\vspace{3mm}

\textit{Department of Physics, University of Toronto, \\60 St. Georges St.,
Toronto, ON M5S 1A7, Canada\\
Department of Chemical and Physical Sciences, University of Toronto,\\
3359 Mississauga Road North, Mississauga, Ontario L5L 1C6, Canada} \\
\vspace{5mm}
\end{center}
\begin{abstract}
New spherically symmetric solution in 4D Einstein--Gauss--Bonnet gravity coupled with nonlinear electrodynamics is obtained. At infinity this
solution has the Reissner--Nordstr\"{o}m behavior of the charged black hole.
The black hole thermodynamics, entropy, shadow, energy emission rate and quasinormal modes of black holes are investigated.
\end{abstract}

\vspace{3mm}
\vspace{3mm}

\section{Introduction}

The heterotic string theory at the low energy limit gives the action including higher order curvature terms \cite{Witten}. Glavan and Lin proposed a new theory of gravity in four dimensions, 4D Einstein--Gauss--Bonnet gravity (4D EGB) \cite{Glavan}, with higher order curvature corrections. The action of the 4D EGB theory consists of the Einstein$-$Hilbert action and the Gauss--Bonnet (GB)
term which is a case of the Lovelock theory. The Lovelock gravity represents the generalization of Einstein's general relativity in higher dimensions that leads to covariant second-order field equations. The Einstein--Gauss--Bonnet gravity in 5D and higher dimensions was studied in \cite{Cvetic}. 4D EGB gravity recently paid much attention \cite{Deser}-\cite{Hennigar}.
Glavan and Lin showed \cite{Glavan} that the GB term, which is a topological invariant before regularization, after
regularization, rescaling the coupling constant, it contributes to the equation of motion.
The authors of \cite{Cai}, \cite{Cai1} found a solution of the semi-classical Einstein equations with conformal anomaly which is also a solution in the 4D EGB gravity. The approach of Glavan and Lin was recently debated in \cite{Tekin}-\cite{Hohmann}. It was shown by \cite{Kobayashi}, \cite{Bonifacio} that solutions in the 4D EGB theory are different from GR solutions as they are due to extra infinitely strongly coupled scalars.
The authors of \cite{Aoki}, \cite{Aoki1}, and \cite{Aoki2} proposed a consistent theory of 4D EGB gravity with two dynamical degrees of freedom that breaks the temporal diffeomorphism invariance in agreement with the Lovelock theorem.
In accordance with the Lovelock theorem \cite{Lovelock}, for a novel 4D theory with two degrees of freedom, the 4D diffeomorphism invariance has to be broken. In the theory of \cite{Aoki}-\cite{Aoki2} the invariance
under the 3D spatial diffeomorphism holds. The authors considered EGB gravity in arbitrary D-dimensions with the Arnowitt--Deser--Misner
decomposition. Then they regularized the Hamiltonian with counter terms where
$D-1$ diffeomorphism invariance holds and taking the limit $D\rightarrow 4$.
It should be noted that the theory of \cite{Aoki}-\cite{Aoki2}, in the spherically-symmetric metrics, represents the solution  that is a solution in the scheme of \cite{Glavan} (see \cite{Lobo1}).
In this work, we obtain a black hole (BH) solution in the 4D EGB gravity coupled with nonlinear electrodynamics (NED) proposed in \cite{Krug2} in the framework of \cite{Aoki}-\cite{Aoki2} theory. Quasinormal modes, deflection angle, shadows of BHs and the Hawking radiation were studied in \cite{Konoplya1}-\cite{Wei1}.
The image of the M87* BH, observed by the Event Horizon Telescope collaboration \cite{Event1}, confirms the existence of BHs in the universe.
The BH shadow is the closed curve which separates capture orbits and scattering orbits. A review on BH shadows see, for example, in \cite{Dokuchaev}.

The paper is organized as follows. In Sec. 2, we find BH spherically symmetric solution in the 4D EGB gravity. It is shown that at infinity we have the Reissner$-$Nordstr\"{o}m behavior of the charged BH. We study the BH thermodynamics in Sec. 3. The Hawking temperature and the heat capacity are calculated showing the possibility of second order phase transitions. The entropy of BHs are obtained which includes the area law and the logarithmic correction. In Sec. 4 the BH shadow is studied. The photon sphere radii, the event horizon radii, and the shadow radii are calculated. We investigate the BH energy emission rate in Sec. 5.  In Sec. 6 quasinormal modes are studied and we obtain the complex frequencies. In Sec. 7 we make a conclusion.

\section{The model}

The action of the EGB gravity in D-dimensions coupled to NED is given by
\begin{equation}
I=\int d^Dx\sqrt{-g}\left[\frac{1}{16\pi G}\left(R+ \alpha{\cal L}_{GB}\right)+{\cal L}_{NED}\right],
\label{1}
\end{equation}
where $\alpha$ has the dimension of (length)$^2$ and the Lagrangian of NED, proposed in \cite{Krug2}, is
\begin{equation}
{\cal L}_{NED} = -\frac{{\cal F}}{\cosh\left(\sqrt[4]{|\beta{\cal F}|}\right)},
 \label{2}
\end{equation}
with the parameter $\beta$ ($\beta\geq 0$) having the dimension of (length)$^4$, ${\cal F}=(1/4)F_{\mu\nu}F^{\mu\nu}=(B^2-E^2)/2$, $F_{\mu\nu}=\partial_\mu A_\nu-\partial_\nu A_\mu$ is the field strength tensor.
The GB Lagrangian reads
\begin{equation}
{\cal L}_{GB}=R^{\mu\nu\alpha\beta}R_{\mu\nu\alpha\beta}-4R^{\mu\nu}R_{\mu\nu}+R^2.
\label{3}
\end{equation}
The variation of action (1) with respect to the metric results in field equations
\begin{equation}
R_{\mu\nu}-\frac{1}{2}g_{\mu\nu}R+\alpha H_{\mu\nu}=-8\pi GT_{\mu\nu},
\label{4}
\end{equation}
where
\begin{equation}
H_{\mu\nu}=2\left(RR_{\mu\nu}-2R_{\mu\alpha}R^\alpha_{~\nu}-2R_{\mu\alpha\nu\beta}R^{\alpha\beta}-
R_{\mu\alpha\beta\gamma}R^{\alpha\beta\gamma}_{~~~\nu}\right)-\frac{1}{2}{\cal L}_{GB}g_{\mu\nu}.
\label{5}
\end{equation}
In the following we consider a magnetic BH with the spherically symmetric field. The static and spherically symmetric metric in $D$ dimension is given by
\begin{equation}
ds^2=-f(r)dt^2+\frac{dr^2}{f(r)}+r^2d\Omega^2_{D-2},
\label{6}
\end{equation}
with $d\Omega^2_{D-2}$ being the line element of the unit $(D - 2)$-dimensional sphere.
Equations (1), (3)-(5) are valid in $D$ dimensions and we will consider rescaled $\alpha$ as $\alpha\rightarrow \alpha/(D-4)$ and then the limit $D\rightarrow 4$.
Taking into account that the electric charge $q_e=0$, ${\cal F}=q^2/(2r^4)$ ($q$ is a magnetic charge), one obtains the magnetic energy density \cite{Krug2}
\begin{equation}\label{7}
  \rho=T_0^{~0}=-{\cal L} = \frac{{\cal F}}{\cosh\left(\sqrt[4]{|\beta{\cal F}|}\right)}=\frac{1}{\beta x^4\cosh\left(1/x\right)},
\end{equation}
where we introduced the dimensionless variable $x=2^{1/4}r/(\beta^{1/4}\sqrt{q})$.
We consider the limit $D \rightarrow 4$ and at $\mu=\nu=t$ field equation (4) gives
\begin{equation}
r(2\alpha f(r)-r^2-2\alpha)f'(r)-(r^2+\alpha f(r)-2\alpha)f(r)+r^2-\alpha=2r^4G\rho.
\label{8}
\end{equation}
Making use of Eq. (7 ) we obtain
\begin{equation}\label{9}
\int_0^r r^2\rho dr=m_M-\frac{2^{1/4}q^{3/2}}{\beta^{1/4}}\arctan\left(\tanh\left(\frac{\beta^{1/4}\sqrt{q}}{2^{5/4}r}\right)\right),
\end{equation}
where the magnetic mass of the black hole reads
\begin{equation}\label{10}
m_M=\int_0^\infty r^2\rho dr=\frac{\pi q^{3/2}}{2^{7/4}\beta^{1/4}}.
\end{equation}
Then the solution to Eq. (8) is
\[
f(r)=1+\frac{r^2}{2\alpha}\left(1\pm\sqrt{1+\frac{8\alpha G}{r^3}(m+h(r)}\right),
\]
\begin{equation}
h(r)=m_M-\frac{2^{1/4}q^{3/2}}{\beta^{1/4}}\arctan\left(\tanh\left(\frac{\beta^{1/4}\sqrt{q}}{2^{5/4}r}\right)\right),
\label{11}
\end{equation}
where $m$ is the Schwarzschild mass (the constant of integration) and $M=m+m_M$ is the total mass of the BH. One can verify that the Weyl tensor for the $D$-dimensional spatial part of the spherically symmetric $D$-dimensional line element (6) vanishes \cite{Lobo1}. As a result, the new solution (11) obtained in the framework of \cite{Glavan} is also a solution for the consistent theory \cite{Aoki}-\cite{Aoki2}.
For Maxwell electrodynamics the energy density is $\rho=q^2/(2r^4)$ and Eq. (8) leads to the metric function obtained in \cite{Fernandes}. In the dimensionless form Eq. (11) becomes
\begin{equation}
f(x)=1+Cx^2\pm C\sqrt{x^4+x(A-Bg(x))},
\label{12}
\end{equation}
where
\[
A=\frac{2^{15/4}(m+m_M)\alpha G}{\beta^{3/4}q^{3/2}},~~~B=\frac{16\alpha G}{\beta},~~~C=\frac{\sqrt{\beta}q}{2\sqrt{2}\alpha},
\]
\begin{equation}\label{13}
 g(x)=\arctan\left(\tanh\left(\frac{1}{2x}\right)\right),
\end{equation}
 We will use the sign minus of the square root in Eqs. (11) and (12) (the negative branch) because in this case the BH is stable and without ghosts \cite{Deser}.
The asymptotic of the metric function $f(r)$ (11), for the negative branch, is given by
\begin{equation}
f(r)=1-\frac{2MG}{r}+\frac{Gq^2}{r^2}+{\cal O}(r^{-3})~~~~r\rightarrow \infty,
\label{14}
\end{equation}
where  the total mass of the BH $M=m+m_M$ includes the Schwarzschild mass $m$ and the electromagnetic mass $m_M$.
% Equation (18) shows that we have the regular BH because $f(r)\rightarrow 1$ as $r\rightarrow 0$.
According to Eq. (14) the Reissner$-$Nordstr\"{o}m behavior of the charged BH holds at infinity.
It is worth noting that the limit $\beta\rightarrow 0$ has be in Eq. (8) before the integration. In this case the solution to Eq. (8) at $\beta=0$ is given by \cite{Fernandes}. The plot of the function (12) is given in Fig. 1.
\begin{figure}[h]
\includegraphics[height=4.0in,width=4.0in]{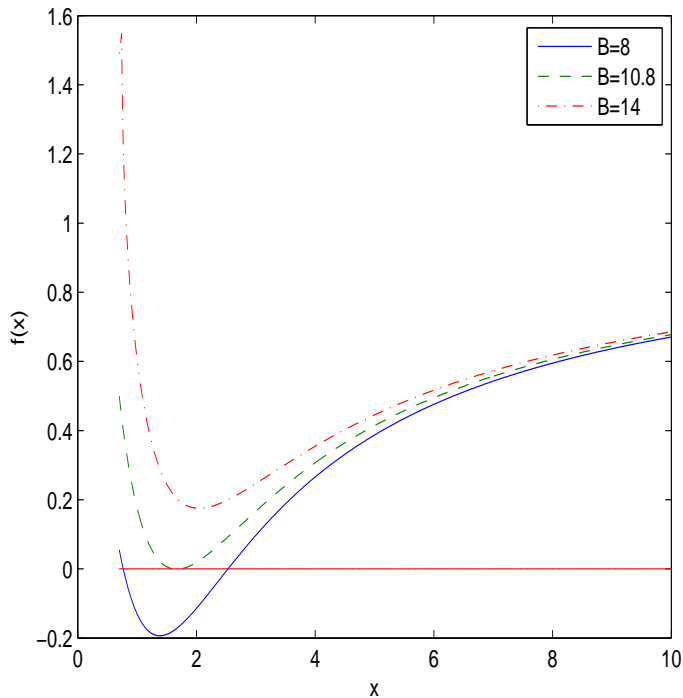}
\caption{\label{fig.1}The plot of the function $f(x)$ for $A=7, C=1$.}
\end{figure}
In accordance with Fig. 1 we have two horizons, one (the extreme) horizon or not horizons depending on the model parameters.

\section{The BH thermodynamics}

Consider the BH thermodynamics and the thermal stability of the BH. The Hawking temperature is given by
\begin{equation}
T_H(r_+)=\frac{f'(r)\mid_{r=r_+}}{4\pi},
\label{15}
\end{equation}
where $r_+$ is the event horizon radius defined by the biggest root of the equation $f(r_h)=0$. Making use of Eqs. (12) and (15), with the variable $x=2^{1/4}r/\sqrt[4]{\beta q^2}$, we obtain the Hawking temperature
\begin{equation}
T_H(x_+)=\frac{2^{1/4}}{4\pi \sqrt[4]{\beta q^2}}\left(\frac{2cx_+^2-1+BC^2x_+^2g'(x_+)}{2x_+(1+cx_+^2)}\right),
\label{16}
\end{equation}
\[
g'(x_+)=-\frac{1}{2x_+^2\cosh^2(1/(2x_+))(\tanh^2(1/(2x_+))+1)},
\]
where we substituted parameter $A$ from equation $f(x_+)=0$.
The plot of the dimensionless function $T_H(x_+)\sqrt[4]{\beta q^2}$ versus $x_+$ is depicted in Fig. 2.
\begin{figure}[h]
\includegraphics[height=4.0in,width=4.0in]{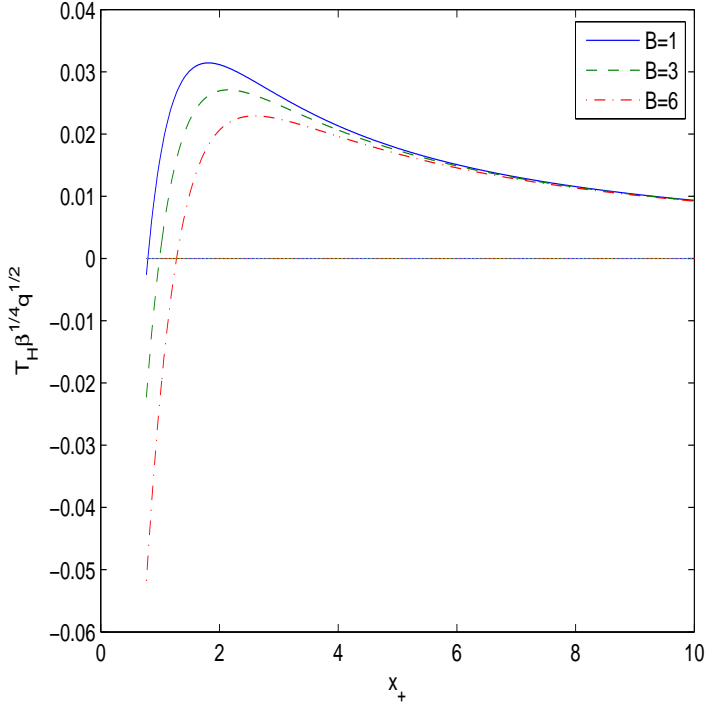}
\caption{\label{fig.2}The plot of the function $T_H(x_+)\sqrt[4]{\beta q^2}$ at $C=1$.}
\end{figure}
According to Fig. 2 the Hawking temperature is positive in some range of $x_+$.
To study the local stability of the BH we calculate the heat capacity making use of the expression
\begin{equation}
C_q(x_+)=T_H\left(\frac{\partial S}{\partial T_H}\right)_q=\frac{\partial M(x_+)}{\partial T_H(x_+)}=\frac{\partial M(x_+)/\partial x_+}{\partial T_H(x_+)/\partial x_+},
\label{17}
\end{equation}
where $M(x_+)$ is the BH gravitational mass depending on the event horizon radius. From equation $f(x_+)=0$ one obtains the BH gravitational
mass
\begin{equation}
M(x_+)=\frac{\beta^{3/4}q^{3/2}}{2^{15/4}\alpha G}\left(\frac{1+2Cx_+^2}{C^2x_+}+Bg(x_+)\right).
 \label{18}
\end{equation}
With the aid of Eqs. (16) and (18) we find
\begin{equation}
\frac{\partial M(x_+)}{\partial x_+}=\frac{\beta^{3/4}q^{3/2}}{2^{15/4}\alpha G}\left(\frac{2Cx_+^2-1}{C^2x_+^2}+Bg'(x_+)\right),
\label{19}
\end{equation}
\[
\frac{\partial T_H(x_+)}{\partial x_+}=\frac{1}{4\pi 2^{3/4}\sqrt[4]{\beta q^2}}\biggl(\frac{5Cx_+^2-2C^2x_+^4+1}{x_+^2(1+Cx_+^2)^2}
\]
\begin{equation}
+\frac{BC^2[g'(x_+)(1-Cx_+^2)+x_+g''(x_+)(1+Cx_+^2)]}{(1+Cx_+^2)^2}\biggr),
\label{20}
\end{equation}
\[
g''(x_+)=\frac{(\tanh^2(1/(2x_+))+1)(2x_+-\tanh(1/(2x_+)))}{2x_+^4\cosh^2(1/(2x_+))(\tanh^2(1/(2x_+))+1)^2}
\]
\[
-\frac{\tanh(1/(2x_+))}{2x_+^4\cosh^4(1/(2x_+))(\tanh^2(1/(2x_+))+1)^2}.
\]
According to Eq. (17) the heat capacity possesses a singularity when the Hawking temperature has an extremum, $\partial T_H(x_+)/\partial x_+=0$.
It follows from Eqs. (16) and (17) that at some point, $x_+=x_1$, the Hawking temperature and heat capacity are zero where a first-order phase transition occurs. In this point $x_1$ the BH remnant with not zero BH mass is formed,  but the Hawking temperature and heat capacity become zero. In the point $x=x_2$, $\partial T_H(x_+)/\partial x_+=0$, the heat capacity has a discontinuity and the second-order phase transition takes place. In the interval $x_2>x_+>x_1$ BHs are locally stable and at $x_+>x_2$ the BH becomes unstable.
By using Eqs. (17), (19) and (20) we represented the heat capacity in Fig. 3.
\begin{figure}[h]
\includegraphics[height=4.0in,width=4.0in]{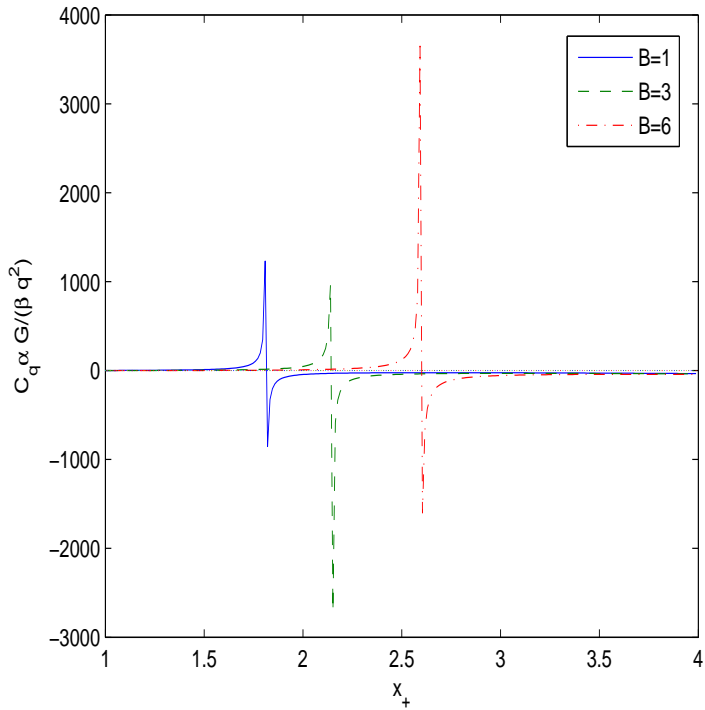}
\caption{\label{fig.3}The plot of the function $C_q(x_+)\alpha G/(\beta q^2)$ at $C=1$. }
\end{figure}
In accordance with Fig. 3 the BH is locally stable in the range $x_2>x_+>x_1$ with the positive Hawking temperature and heat capacity.
The entropy $S$ at the constant charge $q$ could be calculated from the first law of BH thermodynamics $dM(x_+)=T_H(x_+)dS+\phi dq$,
\begin{equation}
S=\int \frac{dM(x_+)}{T_H(x_+)}=\int \frac{1}{T_H(x_+)}\frac{\partial M(x_+)}{\partial x_+}dx_+.
\label{21}
\end{equation}
It should be noted that the entropy in this expression is defined up to a constant of integration.
Making use of Eqs. (16), (19) and (21) we obtain the entropy
\begin{equation}
S=\frac{\pi\beta q^2}{8 C^2\alpha G}\int\frac{1+Cx_+^2}{x_+}dx_+=\frac{\pi r_+^2}{G}+\frac{4\pi\alpha}{G}\ln\left(\frac{\sqrt[4]{2}r_+}{\sqrt[4]{\beta q^2}}\right)+Constant,
\label{22}
\end{equation}
were $Constant$ is the integration constant. One can see the discussion of integration constants in \cite{Medved}. We choose the integration constant as
\begin{equation}
Constant=\frac{2\pi\alpha}{G}\ln\left(\frac{\pi q\sqrt{\beta}}{\sqrt{2}G}\right).
\label{23}
\end{equation}
From Eqs. (22) and (23) we find the BH entropy
\begin{equation}
S=S_0+\frac{2\pi\alpha}{G}\ln\left(S_0\right),
\label{24}
\end{equation}
where $S_0=\pi r_+^2/G$ is the Bekenstein--Hawking entropy. According to Eq. (24) there is the logarithmic correction to area law.
The entropy (24) does not contain the NED parameter $\beta$. The entropy (24) was obtained in 4D EGB gravity coupled to other NED models in
\cite{Kruglov2}-\cite{Kruglov3}. Thus, entropy (24) does not depend on NED is due to GB term in the action and the logarithmic correction vanishes when $\alpha =0$.
At big $r_+$ (event horizon radii) the Bekenstein--Hawking entropy is dominant and for small $r_+$ the logarithmic correction is important.
It is worth noting that at some event horizon radius $r_0$ the entropy vanishes and when $r_+<r_0$ the entropy becomes negative. The negative entropy of BHs was discuss in \cite{Cvetic}.

\section{The shadow of black holes}

The shadow of the BH is due to the light gravitational lensing and is a black circular disk.
The image of the super-massive M87* BH was observed by the Event Horizon Telescope collaboration \cite{Event1}. The shadow of a neutral Schwarzschild BH was investigated in \cite{Synge}. The photons moving in the equatorial plane with $\vartheta=\pi/2$ will be considered.
Making use of the Hamilton$-$Jacobi method, the photon motion in null curves is described by the equation (see, for example, \cite{Kruglov1})
\begin{equation}
H=\frac{1}{2}g^{\mu\nu}p_\mu p_\nu=\frac{1}{2}\left(\frac{L^2}{r^2}-\frac{E^2}{f(r)}+
\frac{\dot{r}^2}{f(r)}\right)=0,
\label{25}
\end{equation}
with $p_\mu$ being the photon momentum, $\dot{r}=\partial H/\partial p_r$, the energy and angular momentum of a photon, which are constants of motion, are defined by $E=-p_t$ and $L=p_\phi$ . Equation (21) can be represented in the form
\begin{equation}
V+\dot{r}^2=0, ~~~V=f(r)\left(\frac{L^2}{r^2}-\frac{E^2}{f(r)}\right).
\label{26}
\end{equation}
The radius of the photon circular orbit  $r_p$ obeys the equation $V(r_p)=V'(r)_{|r=r_p}=0$. From Eq. (26) one obtains
\begin{equation}
\xi\equiv\frac{L}{E}=\frac{r_p}{\sqrt{f(r_p)}},~~~f'(r_p)r_p-2f(r_p)=0,
\label{27}
\end{equation}
with $\xi$ being the impact parameter.
The shadow radius $r_s$ for a distant observer, $r_0\rightarrow \infty$, reads $r_s=r_p/\sqrt{f(r_p)}$.
Note that the impact parameter is $\xi=r_s$. The event horizon radius $r_+$ is the biggest root of the equation $f(r_h)=0$. Making use of Eq. (12) and $f(r_h)=0$ one finds the parameters $A$, $B$ and $C$ versus $x_h$
\[
A=\frac{1+2Cx_h^2+C^2x_hBg(x_h))}{C^2x_h},~~~B=\frac{-1-2Cx_h^2+C^2x_hA}{C^2x_hg(x_h))},
\]
\begin{equation}
C=\frac{x_h^2+\sqrt{x_h^4+x_h(A-Bg(x_h))}}{x_h(A-Bg(x_h))},
\label{28}
\end{equation}
with $x_h=r_h/\sqrt[4]{\beta q^2}$.
The plots of functions (28) are given in Fig. 4.
\begin{figure}[h]
\includegraphics [height=4.0in,width=6.0in]{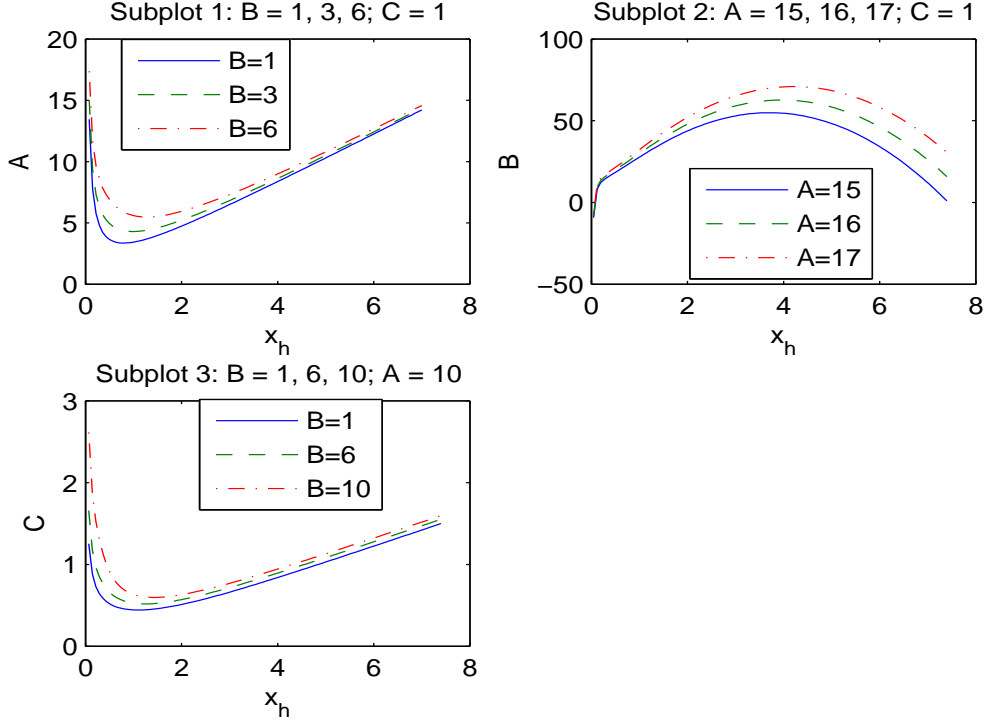}
\caption{\label{fig.4}The plot of the functions $Ax_h)$, $B(x_h)$, $C(x_h)$ }
\end{figure}
According to Fig. 4 (Subplot 1) if parameter $A$ increases, the event horizon radius $x_+$ also increases. Figure 4 (Subplot 2) shows that when parameter $B$ increases, the event horizon radius decreases.
According to Fig. 4 (Subplot 3) if $C$ increasing the event horizon radius $x_+$ also increasing.

In Table 1 we presents the photon sphere radii ($x_p$), the event horizon radii ($x_+$), and the shadow radii ($x_s$) for $A=7$ and $C=1$. The null geodesics radii $x_p$ belong to unstable orbits and correspond to the maximum of the potential $V(r)$ ($V''\leq 0$).
\begin{table}[ht]
\caption{The event horizon, photon sphere and shadow dimensionless radii for A=7, C=1}
\centering
\begin{tabular}{c c c c c c c c c c  c}\\[1ex]
% centered columns
\hline
$B$ & 0.1 & 0.5 & 1  & 2 & 3 & 4 & 5 & 6  \\[0.5ex]
\hline
 $x_+$ & 3.34 & 3.31 & 3.27 & 3.19 & 3.10 & 3.01 & 2.91 & 2.80  \\[0.5ex]
\hline
 $x_p$ & 5.11 & 5.07 & 5.02 & 4.91 & 4.80 & 4.68 & 4.55 & 4.42 \\[0.5ex]
\hline
 $x_s$ & 8.97 & 8.92 & 8.85 & 8.71 & 8.57 & 8.43 & 8.27 & 8.11 \\[0.5ex]
\hline
\end{tabular}
\end{table}
According to Table 1, when the parameter $B$ increasing the shadow radius $x_s$ decreases.
%Because the parameter $c=1$ in Table 1, we find $\alpha=\sqrt{\beta}q_m/2$ and $b=4\alpha G/\beta=2q_mG/\sqrt{\beta}$. Therefore, if the %parameter $b$ is increased (at fixed $a$, $c$, and $q_m$) the parameter $\beta$ decreasing and the BH shadow radius is decreased. In other %words, increasing nonlinearity in NED the BH shadow radius increasing.
Because $x_s>x_+$  the BH shadow radius is given by the radius $r_s=x_s\sqrt[4]{\beta q^2}/2^{1/4}$.

It is worth noting that nonlinear interaction of fields in the framework of NED leads to the self-interaction, and photons propagate along null geodesics of the effective metric \cite{Novello}, \cite{Novello1}. But corrections in radii of photon spheres and impact parameters (due to the self-interaction of electromagnetic fields) are small \cite{Kocherlakota}.

\section{The energy emission rate of black holes}

For the observer at infinity the BH shadow is linked with the high energy absorption cross section
\cite{Wei}, \cite{Belhaj}. The absorption cross-section, at very high energies, oscillates around the photon sphere $\sigma\approx \pi r_s^2$ and the BH energy emission rate is expressed as
\begin{equation}
\frac{d^2E(\omega)}{dtd\omega}=\frac{2\pi^3\omega^3r_s^2}{\exp\left(\omega/T_H(r_+)\right)-1},
\label{29}
\end{equation}
with $\omega$ being the emission frequency. From Eqs. (16) and (29) we obtain the BH energy emission rate in terms of the dimensionless variable $x_+=2^{1/4}r_+/\sqrt[4]{\beta q^2}$
\begin{equation}
\beta^{1/4}\sqrt{q}\frac{d^2E(\omega)}{dtd\omega}=
\frac{2\pi^3\varpi^3x_s^2}{\exp\left(\varpi/\bar{T}_H(x_+)\right)-1},
\label{30}
\end{equation}
where $\bar{T}_H(x_+)=\beta^{1/4}\sqrt{q}T_H(x_+)$ and $\varpi=\beta^{1/4}\sqrt{q}\omega$.
The radiation rate, as a function of the dimensionless emission frequency $\bar{\omega}$  for $C=1$, $A=7$ and $B=0.1, 3, 6$, is plotted in Fig. 5.
\begin{figure}[h]
\includegraphics[height=4.0in,width=4.0in]{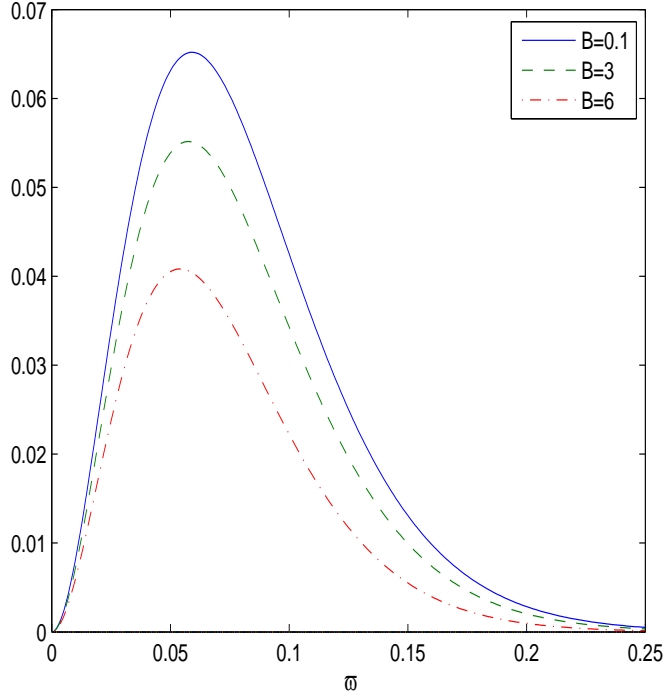}
\caption{\label{fig.6}The plot of the function $\beta^{1/4}\sqrt{q}\frac{d^2E(\omega)}{dtd\omega}$ vs. $\varpi$ for $B=0.1, 3, 6$, $A=7$, $C=1$.}
\end{figure}
According to Fig. 5 we have a peak of the BH energy emission rate. If the parameter $B$ increases, the peak of the energy emission rate becomes smaller and being in the low frequency. At a bigger parameter $B$ the BH possesses a bigger lifetime.

\section{Quasinormal modes}

The information about the stability of BHs under small perturbations can be obtained by studding quasinormal modes (QNMs) which are characterised by complex frequencies $\omega$. The mode is stable when Im~$\omega<0$  otherwise it is unstable. In the eikonal limit Re~$\omega$ is connected with the radius of the BH shadow \cite{Jusufi2}, \cite{Jusufi3}. The perturbations by a scalar massless field around BHs are described by
the effective potential barrier
\begin{equation}
V(r)=f(r)\left(\frac{f'(r)}{r}+\frac{l(l+1)}{r^2}\right),
\label{31}
\end{equation}
where $l$ is the multipole number $l=0,1,2...$.
Equation (31) can be represented as
\begin{equation}
V(x)\sqrt{\beta}q=\sqrt{2}f(x)\left(\frac{f'(x)}{x}+\frac{l(l+1)}{x^2}\right).
\label{32}
\end{equation}
The dimensionless potential $V(x)\sqrt{\beta}q$ is given in Fig. 6 for $A=7$, $B=1$, $C=1$ (Subplot 1) and $l=3,4,5$ and for $A=7$, $C=1$, $l=5$ and $B=1,3,6$ (Subplot 2).
\begin{figure}[h]
\includegraphics{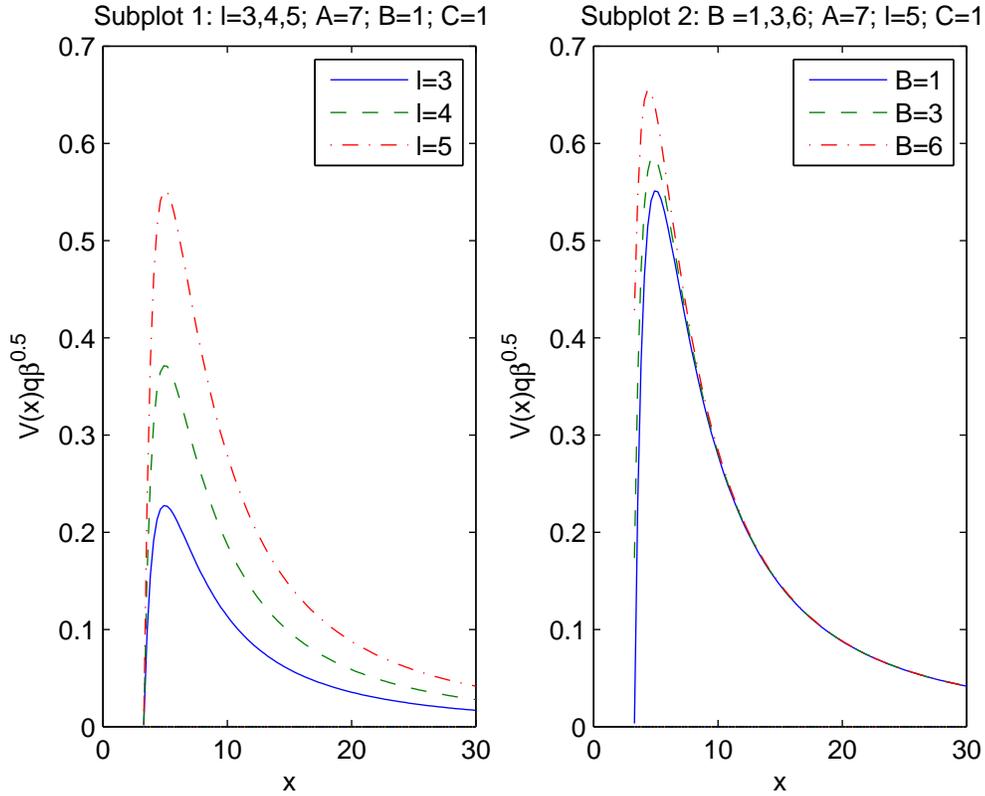}
\caption{\label{fig.6}The plot of the function $V(x)\sqrt{\beta} q$ for $A=7$, $C=1$.}
\end{figure}
Figure 6, Subplot l, shows that the potential barriers of effective potentials have the maxima. When the $l$ increases the height of the potential increases. According to Fig. 6, Subplot 2, if the parameter $B$ increases the height of the potential increases.
The quasinormal frequencies can be found by \cite{Jusufi2}, \cite{Jusufi3}
\begin{equation}
\mbox{Re}~\omega=\frac{l}{r_s}=\frac{l\sqrt{f(r_p)}}{r_p},~~~~\mbox{Im}~\omega=-\frac{2n+1}{2\sqrt{2}r_s}\sqrt{2f(r_p)-r_p^2f''(r_p)},
\label{33}
\end{equation}
with $r_s$ being the BH shadow radius, $r_p$ is the BH photon sphere radius, and $n=0,1,2,...$ is the overtone number.
The frequencies, depending on parameter $B$ (at $A=7$, $C=1$, $n=1$, $l=5$), are represented in Table 2.
\begin{table}[ht]
\caption{The real and the imaginary parts of the frequencies vs the parameter $B$ at $n=1$, $l=5$,  $A=7$, $C=1$}
\centering
\begin{tabular}{c c c c c c c c c c  c}\\[1ex]
% centered columns
\hline
$B$  & 0.1 & 0.5 &  1 & 2 & 3 & 4 & 5 & 6  \\[0.5ex]
\hline
 $\sqrt[4]{\beta q^2}\mbox{Re}~\omega$  & 0.557 & 0.561 & 0.565 & 0.574 & 0.583 & 0.593 & 0.605 & 0.617  \\[0.5ex]
\hline
 $-\sqrt[4]{\beta q^2}\mbox{Im}~\omega$ & 0.3212 & 0.3215 & 0.3221 & 0.3229 & 0.3234 & 0.3232 & 0.3230 & 0.3220  \\[0.5ex]
\hline
\end{tabular}
\end{table}
The modes are stable (the real part represents the frequency of oscillations) because the imaginary parts of the frequencies in Table 2 are negative. Table 2 shows that when the parameter $B$ increases the real part of the frequency $\sqrt[4]{\beta q^2}\mbox{Re}~\omega$ increases, and the absolute value of the imaginary part of the frequency $\mid\sqrt[4]{\beta q^2}\mbox{Im}~\omega\mid$ increases. Therefore, when the parameter $B$ is increased the scalar perturbations oscillate with greater frequency and decay fast.

\section{Conclusion}

We have obtained the exact spherically symmetric and magnetized BH solution in 4D EGB gravity coupled with NED. The thermodynamics and the thermal stability of  magnetically charged BHs were studied  by calculating the Hawking temperature and the heat capacity. The phase transitions take place in the points where the Hawking temperature possesses the extremum. It is shown that BHs are thermodynamically stable at some interval of event horizon radii when the heat capacity and the Hawking temperature are positive. The heat capacity possesses a singularity in some event horizon radii where the second-order phase transitions occur. The entropy of BHs is calculated including the Hawking entropy and the logarithmic correction. The photon sphere radii, the event horizon radii, and the shadow radii are calculated. We show that with increasing the model parameter $B$ the BH energy emission rate decreases and, as a result, the BH has a bigger lifetime. The quasinormal modes are investigated and it is shown that increasing the parameter $B$ the scalar perturbations oscillate with greater frequency and decay fast. It is worth noting that other solutions in 4D EGB gravity coupled to some NED were obtained in \cite{Kruglov2}-\cite{Kruglov3}. It is of interest to study solutions of BHs in 4D EGB gravity coupled to different NED because astrophysical characteristics depend on them.


\begin{thebibliography}{99}

\bibitem{Witten} D. J. Gross and E. Witten, Nucl. Phys. B \textbf{277}, 1 (1986); D. J. Gross and J. H. Sloan, Nucl. Phys. B \textbf{291}, 41 (1987); R. R. Metsaev and A. A. Tseytlin, Phys. Lett. B \textbf{191}, 354 (1987); B. Zwiebach, Phys. Lett. B \textbf{156}, 315 (1985); R. R. Metsaev and A. A. Tseytlin, Nucl. Phys. B \textbf{293}, 385 (1987).
\bibitem{Glavan} D. Glavan and C. Lin, Phys. Rev. Lett. \textbf{124}, 081301 (2020).
\bibitem{Cvetic} M. Cvetic, S. I. Nojiri, and S. D. Odintsov, Nucl. Phys. B \textbf{628}, 295 (2002).
%Black hole thermodynamics and negative entropy in de Sitter and anti-de Sitter Einstein-Gauss-Bonnet gravity, e-Print: hep-th/0112045 [hep-th]
\bibitem{Deser}D. G. Boulware and S. Deser, Phys. Rev. Lett. \textbf{55}, 2656 (1985); J. T. Wheeler, Nucl. Phys. B \textbf{268}, 737 (1986);  R. C. Myers and J. Z. Simon, Phys. Rev. D \textbf{38}, 2434 (1988).
\bibitem{Lovelock}D. Lovelock, J. Math. Phys. \textbf{12}, 498 (1971).
\bibitem{Cai}R. G. Cai, L. M. Cao, and N. Ohta, JHEP \textbf{1004}, 082 (2010) [arXiv:0911.4379 [hep-th]].
\bibitem{Cai1} R.-G. Cai, Phys. Lett. B \textbf{733}, 183 (2014) [arXiv:1405.1246 [hep-th]].
\bibitem{Cognola} G. Cognola, R. Myrzakulov, L. Sebastiani, and S. Zerbini, Phys. Rev. D \textbf{88}, 024006   (2013) [arXiv:1304.1878 [gr-qc]].
\bibitem{Fernandes} P. G. S. Fernandes, Phys. Lett. B \textbf{805} 135468 (2020) [arXiv:2003.05491 [gr-qc]].
%\bibitem{[7]}D. V. Singh, S. G. Ghosh and S. D. Maharaj, [arXiv:2003.14136 [gr-qc]].
\bibitem{Jusufi}K. Jusufi, Ann. Phys. \textbf{421}, 168285 (2020) [arXiv:2005.00360 [gr-qc]].
\bibitem{Ghosh} S. G. Ghosh, D. V. Singh, R. Kumar, and S. D. Maharaj, Ann. Phys. \textbf{424}, 168347 (2021) [arXiv:2006.00594 [gr-qc]].
%Phase transition of AdS black holes in 4D EGB gravity coupled to nonlinear electrodynamics 
\bibitem{Ghosh1} S. G. Ghosh and S. D. Maharaj, Phys. Dark Univ. \textbf{30}, 100687 (2020) [arXiv:2003.09841 [gr-qc]].
 % arXiv:2003.09841 [gr-qc] e-Print: .
\bibitem{Ghosh2} R. Kumar and S. G. Ghosh, JCAP \textbf{07}, 053 (2020) [arXiv:2003.08927 [gr-qc]].
\bibitem{Jin} X. H. Jin, Y. X. Gao, and D. J. Liu, Int. J. Mod. Phys. D \textbf{29}, 2050065 (2020) [arXiv:2004.02261 [gr-qc]].
\bibitem{Jusufi1} K. Jusufi, A. Banerjee, and S. G. Ghosh, Eur. Phys. J. C \textbf{80}, 698 (2020).
\bibitem{Guo} M. Guo and P. Li, Eur. Phys. J. C \textbf{80}, 588 (2020) [arXiv:2003.02523 [gr-qc]].
%\bibitem{[26]}C. Zhang, P. Li and M. Guo, [arXiv:2003.13068 [hep-th]].
\bibitem{Zhang} C. Zhang, S. Zhang, P. Li, and M. Guo, JHEP \textbf{08}, 105 (2020) [arXiv:2004.03141[gr-qc]].
\bibitem{Odintsov} S. Odintsov, V. Oikonomou, and F. Fronimos, Nucl. Phys. B \textbf{958}, 115135 (2020) [arXiv:2003.13724 [gr-qc]].
\bibitem{Ai} W. Ai, Commun. Theor. Phys. \textbf{72}, 095402 (2020) [arXiv:2004.02858 [gr-qc]].
\bibitem{Fernandes1} P. G. Fernandes, P. Carrilho, T. Clifton, and D. J. Mulryne, Phys. Rev. D \textbf{102}, 024025 (2020).
\bibitem{Hennigar} R. A. Hennigar, D. Kubiznak, R. B. Mann, and C. Pollack, JHEP \textbf{2020}, 27 (2020) [arXiv:2004.09472 [gr-qc]].
%\bibitem{[44]} R.B. Mann, S.F. Ross, Class. Quant. Grav.10:1405-1408,1993
\bibitem{Tekin}M. Gurses, T. C. Sisman, and B. Tekin, Phys. Rev. Lett. \textbf{125},  149001 (2020) [arXiv:2009.13508 [gr-qc]].
%Comment on "Einstein-Gauss-Bonnet Gravity in 4-Dimensional Space-Time'' e-Print: 
\bibitem{Tekin1} M. Gurses, T. C. Sisman, and B. Tekin, Eur. Phys. J. C \textbf{80}, 647 (2020) [arXiv:2004.03390 [gr-qc]].
\bibitem{Mahapatra} S. Mahapatra, Eur. Phys. J. C \textbf{80}, 992 (2020).
%A note on the total action of 4D Gauss$-$Bonnet theory, arXiv:2004.09214 [gr-qc].
%\bibitem{Tian} S. X. Tian, Z.-H. Zhu, Non-full equivalence of the four-dimensional Einstein-Gauss-Bonnet gravity and Horndeksi gravity for %Bianchi type I metric, arXiv:2004.09954 [gr-qc].
\bibitem{Arrechea} J. Arrechea, A. Delhom, and A. Jiménez-Cano, Chin. Phys. C \textbf{45}, 013107 (2021) [arXiv:2004.12998 [gr-qc]].
%Inconsistencies in four-dimensional Einstein-Gauss-Bonnet gravity gravity .
\bibitem{Arrechea}J. Arrechea, A. Delhom, and A. Jiménez-Cano, Phys. Rev. Lett. \textbf{125}, 149002 (2020) [arXiv:2009.10715 [gr-qc]].
%Comment on “Einstein-Gauss-Bonnet Gravity in Four-Dimensional Spacetime” 
\bibitem{Hohmann} M, Hohmann and C. Pfeifer,  Eur. Phys. J. Plus \textbf{136}, 180 (2021) [arXiv:2009.05459].
%Canonical variational completion and 4D Einstein$-$Gauss$-$Bonnet gravity, .
\bibitem{Kobayashi} T. Kobayashi, JCAP \textbf{07}, 013 (2020) [arXiv:2003.12771 [gr-qc]].
%Effective scalar-tensor description of regularized Lovelock gravity in four dimensions 
\bibitem{Bonifacio} J. Bonifacio, K. Hinterbichler, and L. A. Johnson, Phys. Rev. D \textbf{102}, 024029 (2020) [arXiv:2004.10716 [hep-th]].
%Amplitudes and 4D Gauss-Bonnet Theory • e-Print: 
\bibitem{Aoki} K. Aoki, M. A. Gorji, and S. Mukohyama, Phys. Lett. B \textbf{810}, 135843 (2020) [arXiv:2005.03859 [gr-qc]]
\bibitem{Aoki1} K. Aoki, M. A. Gorji, and S. Mukohyama, JCAP \textbf{2009}, 014 (2020) [arXiv:2005.08428].
\bibitem{Aoki2} K. Aoki, M. A. Gorji, S. Mizuno and S. Mukohyama, JCAP \textbf{2101}, 054 (2021) [arXiv:2010.03973 [gr-qc]].
\bibitem{Lobo1} K. Jafarzade, M. K. Zangeneh, F. S. N. Lobo, JCAP \textbf{04}, 008 (2021) [arXiv:2010.05755 [gr-qc]].
%.Shadow, deflection angle and quasinormal modes of Born-Infeld charged black holes
\bibitem{Krug2} S. I. Kruglov, Int. J. Mod. Phys. A \textbf{33}, 1850023 (2018) [arXiv:1803.02191].
% "\textit{Magnetically charged black hole in frameworkof nonlinear electrodynamics model}" .
\bibitem{Konoplya1} R. A. Konoplya and  A. F. Zinhailo, Eur. Phys. J. C \textbf{80}, 1049 (2020).
\bibitem{Konoplya2} R. A. Konoplya and  A. F. Zinhailo, Phys. Lett. B \textbf{807}, 135607 (2020).
\bibitem{Belhaj} A. Belhaj, M. Benali, A. El Balali, H. El Moumni, and S. E. Ennadifi,  Class. Quant. Grav. \textbf{37}, 215004 (2020) [arXiv:2006.01078].
%Deflection Angle and Shadow Behaviors of Quintessential Black Holes in arbitrary Dimensions,.
\bibitem{Konoplya3} R. A. Konoplya and Z. Stuchlik, Phys. Lett. B \textbf{771}, 597 (2017) [arXiv:1705.05928].
%Are eikonal quasinormal modes linked to the unstable circular null geodesics?, .
\bibitem{Stefanov} I. Z. Stefanov, S. S. Yazadjiev, and G. G. Gyulchev, Phys. Rev. Lett. \textbf{104}, 251103 (2010) [arXiv:1003.1609].
%Connection between black-hole quasinormal modes and lensing in the strong deflection limit,, .
\bibitem{Guo} Y. Guo and Y. G. Miao, Phys. Rev. D \textbf{102}, 084057 (2020) [arXiv:2007.08227].
%Null geodesics, quasinormal modes and the correspondence with shadows in high-dimensional Einstein-Yang-Mills spacetimes,, .
%\bibitem{Lan} C. Lan, Y. G. Miao and H. Yang, Quasinormal Modes and Thermodynamics of Regular Black Holes, arXiv:2008.04609.
\bibitem{Wei1} S. W. Wei and Y. X. Liu, Chin. Phys. C \textbf{44}, 115103 (2020).
\bibitem{Event1} Event Horizon Telescope collaboration, K. Akiyama et al., Astrophys. J. \textbf{875}, L5 (2019).
%L1; ibid L2; ibid L3; ibid L4; ibid L5; ibid L6.
\bibitem{Dokuchaev} V. I. Dokuchaev and N. O. Nazarova, Usp. Fiz. Nauk \textbf{190}, 627 (2020), Phys. Usp. \textbf{63}, 583 (2020) [arXiv:1911.07695].
%Silhouettes of invisible black holes, .
\bibitem{Medved} A. J. M. Medved and E. C. Vagenas,	Phys. Rev. D \textbf{70},  124021 (2004) [arXiv:hep-th/0411022].
%When conceptual worlds collide: The GUP and the BH entropy, 
\bibitem{Kruglov2}S.I. Kruglov, Ann. Phys. \textbf{428}, 168449 (2021) [arXiv: 2104.08099 [gr-qc]].
%Einstein − Gauss − Bonnet gravity with nonlinear electrodynamics
\bibitem{Kruglov4}S.I. Kruglov, Symmetry \textbf{13}, 944 (2021).
%Einstein–Gauss–Bonnet Gravity with Nonlinear Electrodynamics: Entropy, Energy Emission, Quasinormal Modes and Deflection Angle
\bibitem{Kruglov3}S.I. Kruglov, EPL \textbf{133}, 6 (2021) [arXiv:2106.00586].
% Einstein-Gauss-Bonnet gravity with rational nonlinear electrodynamics e-Print: 2106.00586 [physics.gen-ph]
\bibitem{Synge} J. L. Synge, Mon. Not. Roy. Astron. Soc. \textbf{131}, 463 (1966).
% –466.
\bibitem{Kruglov1} S. I. Kruglov, Symmetry \textbf{13}, 204 (2021).
\bibitem{Novello}    M. Novello, V. A. De Lorenci, J. M. Salim, and R. Klippert, Phys. Rev. D \textbf{61}, 045001 (2000).
\bibitem{Novello1} M. Novello, S. E. Perez Bergliaffa, and J. M. Salim, Class. Quant. Grav. \textbf{17}, 3821 (2000) [arXiv:gr-qc/0003052].
    %``Singularities in general relativity coupled to nonlinear electrodynamics", .
%\bibitem{Quiros} R. Garcia-Salcedo, T. Gonzalez, and I. Quiros, Phys. Rev. D \textbf{89}, 084047 (2014).
%M. Zhang and M. Guo, Eur. Phys. J. C \textbf{80},  790 (2020). arXiv:1909.07033 [gr-qc].
\bibitem{Kocherlakota}P. Kocherlakota and L. Rezzolla, Phys. Rev. D, \textbf{6}, 064058 (2020) [arXiv:2007.15593 [gr-qc]].
\bibitem{Wei} S. W. Wei and Y. X. Liu, JCAP 11, 063 (2013) [arXiv:1311.4251].
% Observing the shadow of Einstein-Maxwell-Dilaton-Axion black hole, .
\bibitem{Belhaj} A. Belhaj, M. Benali, A. El Balali, H. El Moumni, and
S. E. Ennadifi, Class. Quant. Grav. \textbf{37}, 215004 (2020) [arXiv:2006.01078].
\bibitem{Jusufi2} K. Jusufi, Phys. Rev. D \textbf{101}, 084055 (2020) [arXiv:1912.13320].
%Quasinormal Modes of Black Holes Surrounded by Dark Matter and Their Connection with the Shadow Radius, .
\bibitem{Jusufi3} K. Jusufi, Phys. Rev. D \textbf{101}, 124063 (2020) [arXiv:2004.04664].
% Connection Between the Shadow Radius and Quasinormal Modes in Rotating Spacetimes.

\end{thebibliography}
\end{document}